\documentclass[12pt]{article}

\usepackage[utf8]{inputenc}
\usepackage[T1]{fontenc}
\usepackage{amssymb}
\usepackage[english]{babel}
\usepackage{amsmath,amssymb,bm}
\usepackage{graphicx}
\usepackage{hyperref}
\usepackage{verbatim}
\usepackage[linesnumbered,lined,boxed,commentsnumbered,ruled]{algorithm2e}
\SetKwComment{Comment}{}{}
\usepackage{algpseudocode}
\usepackage{listings}
\usepackage{bm}

\newtheorem{definition}{Definition}[section]
\newtheorem{example}{Example}[section]
 
\usepackage{defs}
\usepackage{fca}

\begin{document}


\def\taularesum{
\begin{table}[ht]
\centering
\resizebox{\textwidth}{!}{
\begin{tabular}{ |l|l||l|l| }
\hline 
\multicolumn{2}{|c||}{Semantical} & \multicolumn{2}{|c|}{Syntactical} \\
\hline \hline
Horn Clauses / Implications 
& \cite{Baixeries2003SIAM}
&  
&  
\\
Functional Dependencies 
& \cite{Baixeries2012CLA},\cite{Baixeries2004MML},\cite{Baixeries2014AMAI}
&   
& Armstrong Dependencies 
\\
Similarity Dependencies 
& \cite{Baixeries2013CLA}, \cite{Baixeries2018DAM}
&  
& 
\\
Order-like Dependencies 
& \cite{Baixeries2016CLA}
&  
& 
\\
\hline 
\hline
MVD Clauses 
& \cite{Baixeries2007TESI}, \cite{Baixeries2005CLA}, \cite{Baixeries2006ICFCA}
& 
&  
\\
Degenerated MVD 
&  \cite{Baixeries2005ICFCA}, \cite{Baixeries2005CLA}, \cite{Baixeries2006ICFCA}
& \cite{Baixeries2008ICFCA}, \cite{Baixeries2011CLA}
& Symmetric Dependencies 
\\
Multivalued Dependencies 
& \cite{Baixeries2005ICCS}, \cite{Baixeries2005CLA}, \cite{Baixeries2006ICFCA}
&  
&  
\\
\hline 
\hline
Acyclic Join Dependencies 
& \cite{Baixeries2017} 
& \cite{Baixeries2019ICFCA} 
& Acyclic Hypergraphs \\
\hline
\end{tabular}}
\label{taularesum}
\caption{Table of references (non exhaustive) of FCA/lattice-based characterization of
database constraints.}
\end{table}
}

\title{Database Dependencies and Formal Concept Analysis}


\author{Jaume Baixeries \\
Computer Science Department. \\ Universitat Politècnica de Catalunya \\
Jordi Girona, 1-3, Barcelona, Catalonia}

\maketitle    

\begin{abstract}

This is an account of the characterization of database dependencies with Formal Concept Analysis.


\end{abstract}



\section{Introduction}
\label{sec:introduction}

\taularesum

This is a review of the results concerning the characterization of various types of database dependencies in the Relational Database Model (RDBS) using Formal Concept Analysis (FCA).
In order to explain what a characterization of a type of dependency is, we should define what a dependency (from now on we assume dependencies in the RDBM) is.
A general definition can be found in \cite{KANELLAKIS19901073}:

\begin{quote}
Dependencies, in general, are semantically meaningful and syntactically restricted sentences of the predicate calculus that must be satisfied by any "legal" database. \textbf{Paris C. Kanellakis}
\end{quote}

This definition is a way too generic, but precisely this sort of ambiguity or generality (depending on the point of view of the reader) allows to embrace all different sort of dependencies that are later explained in \cite{KANELLAKIS19901073}.

Here, we dissect this definition, and mention three relevant facts about a dependency:

\begin{enumerate}
    \item It is a structure with a \textbf{syntax} (\textit{syntactically restricted sentences of the predicate calculus}).
    \item It is a structure with \textbf{semantics} (\textit{semantically meaningful sentences of the predicate calculus}).
    \item It is a \textbf{restriction} \textit{that must be satisfied by any "legal" database}.
\end{enumerate}

Therefore, we like to say that a dependency is syntax + semantic (as almost everything in computer science) that can be considered both a restriction and a description. It can be considered a \textbf{restriction}, because it is a sentence of the predicate calculus that must apply to a \textit{legal} database (we guess that here legal means \textit{according to some rules}, precisely defined by these dependencies). In fact, this point of view assumes that, given a database, the dependencies act as an oracle that decides whether this database is "legal" or not (we do not discuss what may happen when a database is not "legal": it can be modified in order to become legal, or it could be discarded). This is the usage that most database practitioners give to the dependencies. But it can also be seen as a \textbf{description} of a database. By this we mean that we can deduce what dependencies hold in a database, not in order to change or modify it, but just to have a description of what constraints hold in that database. This would be the point of view of a data analyst.

Regardless of the use that we can give to dependencies, we remark the two-fold nature of a dependency: syntax + semantics, because this duality will fit beautifully to FCA.

This general definition implies that many different types of dependencies may exist, and in fact, if we follow \cite{KANELLAKIS19901073} we will see that there are many different types of them, that can be classified hierarchically. Some of those dependencies have become more popular and useful (or used) than others.

Once we have presented a definition of a dependency, and we have discussed it, we need to define what we mean by the characterization of a dependency, or, rather, a set of dependencies with FCA.

\section{Functional Dependencies}

The characterization of Functional Dependencies (FDs) with FCA was performed from a semantic point of view. The first results relating FD's with FCA were in \cite{DBLP:books/daglib/0095956}, although we can trace them back to \cite{DBLP:journals/jacm/SagivDPF81}. The method used in \cite{DBLP:books/daglib/0095956} (Section 2.4) consists in creating a formal context for a database:

\begin{proposition}[Proposition 28 in \cite{DBLP:books/daglib/0095956}]
\label{prop01}

Let $DB = (T,\atrib)$ be a database, and let $\powerset_2(T)$ the set of sets of two tuples of $DB$. A functional dependency $\fd{X}{Y}$ holds in $DB$ if and only if the implication $\fd{X}{Y}$ holds in the context

\[ \context = (\powerset_2(T), \atrib, I) \]

where 

\[ (t_i, t_j) I (a) \iif t_i(a) = t_j(a) \]

\end{proposition}

This result is called \textbf{binarization} or \textbf{scaling} in FCA parlance. 
This process consists in creating a set of binary tuples (called models in the realm of logic) 
comparing all tuples in a table pairwise.

The first novel result concerning the characterization of FD's with FCA was presented in 
\cite{Baixeries2012CLA} and later extended in \cite{Baixeries2014AMAI}.
This characterization is presented in terms of pattern structures.

\begin{definition}
\label{def01}

Let $DB = (T,\atrib)$ be a database, we define the following pattern structure:

\[ 
(\atrib,(D,\sqcap),\delta) 
\]

where $(D,\sqcap)$ is the set of partitions over $T$ provided 
with the partition intersection operation $\sqcap$,
and $\delta: \atrib \mapsto D$, which is the tightest partition
of $T$ such that $t_i, t_j$ are in the same class in $\delta(a)$
if and only if $t_i(a) = t_j(a)$.

\end{definition}

This pattern structure is based in the lattice $D$ of partitions on the tuples $T$.
The set of partitions is a lattice \cite{Gratzer}, it can be considered a set of pairs
that conform an equivalence class. The intersection of two equivalence classes
(seen as a set of pairs) is also an equivalence class.
The function $\delta$ yields the partition of $T$ \textbf{induced} by an attribute $a$.
This is the idea of many algorithms that deal with the computation of FDS in the RDBM,
for instance, it is the central idea of the algorithm TANE \cite{TANE}.
The characterization is the following:

\begin{proposition}[Proposition 3 in \cite{Baixeries2014AMAI}]
\label{prop02}

A functional dependency $\ad{X}{Y}$ holds in a table $T$ if and only if:
$\{X\}^{\square} = \{XY\}^{\square}$ in the partition pattern structure $(M,(D,\sqcap),\delta)$.

\end{proposition}

This proposition states that the functional dependency $\ad{X}{Y}$ holds in a table $T$ if and only if
the partition of $T$ induced by $X$ is the same as that induced by $XY$
(we remind that \textit{induced} is equivalent to \textit{tightest induced}).
This is exactly what is described in Proposition \ref{prop01}.
In order to understand this, we need to see that 
that stating that the implication $\fd{X}{Y}$ holds in that context, is equivalent to
saying that $X = XY'$, and also, that $X'$ is a set of pairs of tuples of $T$.
This set of pairs is an equivalence class, more precisely the tightest partition induced by $X$.

\section{Multivalued Dependencies}

The first result appears in \cite{Baixeries2005ICCS}, where a characterization in terms of
partitions of attributes and sets of sets of tuples is presented.

\begin{proposition}[Proposition 3 in \cite{Baixeries2005ICCS}]
Let $DB = (T,\atrib)$ be a database and
let $P = \particio{P_1 | P_2, \dots, P_n}$ a partition of $\atrib$.
The partition $P$ matches a set of tuples $C \subseteq T$ if and only if
$C = \Pi_{P_1}(C) \times \Pi_{P_2}(C) \times \dots \times \Pi_{P_n}(C)$.
\end{proposition}

From this relation, it is easy to define a pair of functions from one set
to another.

\begin{definition}
Operator $\phi$: $\particioatrib \rightarrow \Pi$. 
This operator receives a partition of
the set of attributes $\atrib$: $P = \particio{P_1, \dots, P_n}$ 
and returns the set of all the classes of tuples $\{C_1, \dots, C_m\}$
matched by $P$. 
\end{definition}

\begin{definition}
Operator $\psi$: $\Pi \rightarrow \particioatrib$.
This operator receives a set of classes of
tuples $\{C_1, \dots, C_m\}$
and returns a partition of the set of attributes $P$
that is the finest partition of attributes that
matches that set of classes. 
\end{definition}

It is important to note the fact that the partition that is returned must be the
finest possible (a trivial partition would be that in which all attributes
fall in the same equivalence class).
It is the same with the set of all classes (this is, sets) of tuples.
It must be the set that contains the largest possible sets, otherwise
a trivial set of sets (for instance, that which contains only one tuple per set)
would fulfil the condition.
In fact, there is still more to be discussed here, which is the fact that 
it only needs to return the largest classes, this is, that if a class
is matched by a particular partition of attributes but it is contained 
by a still different larger class that is also matched by that same partition, 
this class is not to be returned.
This condition is not necessary but avoids a certain amount of redundancy.
In fact, the set of classes returned is a tolerance relation over $T$.

\begin{proposition}
The operators $\phi$, $\psi$ are a Galois connection.
\label{prop1}
\end{proposition}

\begin{proposition}
The compositions of the operators $\phi, \psi$: 
$\Gamma = \psi.\phi$ and $\Gamma' = \phi.\psi$
are closure operators. 
\end{proposition}

One relevant point from now on is that we assume that MVDs are of the form
$\mvd{X}{Y_1 \mid Y_2 \mid \dots \mid Y_m}$ (they are called generalized MVDs).

The main result is:

\begin{theorem}[Theorem 1 in \cite{Baixeries2005ICCS}]
A (generalized) MVD $\mvd{X}{Y_1}{Y_2 \mid \dots \mid Y_m}$ holds in $DB = (T,\atrib)$
if and only if 
$\Gamma(\particio{X_1 \mid \dots \mid X_n \mid Y})$ = 
$\Gamma(\particio{X_1 \mid \dots \mid X_n \mid Y_1 \mid Y_2 \mid \dots \mid Y_m})$
\end{theorem}

Some other results concerning the characterization of \MVD appears in \cite{Baixeries2005CLA},
but no explicit characterizations in terms of FCA proper.
Concerning \MVD, this paper presents a way to compute a lattice of partitions of attributes to characterize \DMVD,
and show that the lattice of partitions of attributes that characterizes \MVD is a subset of it. Let us describe the method to compute the former lattice and then, we come back to \MVD.

\begin{definition}
Let $\bin(T)$ be the binarization of a database $DB = (T,\atrib)$.
This is, 
\[
\bin(T) = \conjunt{\tupla{t_i(x_1) = t_j(x_1), \dots, t_i(x_n) = t_j(x_n)} \mid t_i, t_j \in T, x_i \in \atrib}
\]
\end{definition}

\begin{definition}

Let $DB = (T,\atrib)$ such that $n = |\atrib|$, and let $t = \conjunt{0,1}^n$.
The function $\parti: \conjunt{0,1}^n \mapsto \particions$ generates a partition out of a binary tuple.
This is, 
\[
\parti(t) = \conjunt{(x_i,x_j) \mid x_i,x_j \in \atrib \text{ and } t(x_i) = 0}
\cup \conjunt{(x_i,x_i) \mid x_i \in \atrib}
\]

(we present a partition as an equivalence class).
For instance, if $t = \tupla{11001}$ and $\conjunt{a,b,c,d,e}$, then
$\parti(t) = \particio{a \mid b \mid cd \mid e}$.
\end{definition}

Finally, we define the closure of a set $S$ by the operation $\wedge$

\begin{example}
Let $S$ be a set with the total operator $\wedge: S \mapsto S$,
then $\tancameet{S}$ is the closure under $\wedge$ of $S$.
\end{example}

The result in \cite{Baixeries2005CLA} states that the lattice of partitions of attributes that characterizes the \DMVD of a database $DB = (T,\atrib)$ is $\tancameet{\parti(\bin(T))}$.

The main result concerning \MVD is that the lattice that charectarizes the \MVD that hold is a subset of this lattice.

\begin{proposition}[Proposition 3 in \cite{Baixeries2005CLA}]
Let $r$ be a relation, and let $\reticledmvd(r)$ and 
$\reticlemvd(r)$ the lattices that characterize the
DMVDs and MVDs respectively that hold in $r$.
Then, $\reticlemvd(r) \subseteq \reticledmvd(r)$.
\label{propinclusio}
\end{proposition}

The results for the characterization of \MVD in \cite{Baixeries2006ICFCA} continue with this previous definition. Basically, they define a Galois connection between the set of partitions of $\atrib$ and the set of antichains of $\lesparts(T)$, this is, the largest possible sets in $\lesparts(\lesparts(T))$ that fulfil the required condition.

It is in \cite{Baixeries2007TESI} when, finally, a formal context between the set of pairs of tuples of a database and the set of pairs of attributes is defined.

This result also appeared in the characterization of \AJD (a generalization of \MVD)
in \cite{Baixeries2017}.

The evolution has been the following:

\begin{center}

\begin{tabular}{c|c|c}
$\atrib$     &  $T$ & \\
\hline \hline
$\particioatrib$  & $\lesparts(\lesparts(T))$ & \cite{Baixeries2005ICCS} \\
$\particioatrib$  & $\antichains(\lesparts(T))$ & \cite{Baixeries2006ICFCA} \\
$\splits(\atrib)$  & $\pairs(T)$ & \cite{Baixeries2007TESI} \\
\end{tabular}
   
\end{center}


\bibliographystyle{plain}
\bibliography{all_material}

\end{document}